\documentclass[showpacs,aps,pra,superscriptaddress,longbibliography]{revtex4-1}
\usepackage{geometry}
\usepackage{amsfonts}
\usepackage{graphicx}
\usepackage{breakcites}
\usepackage{amsmath}
\usepackage{hyperref}
\usepackage{cleveref}
\usepackage{mathtools}
\usepackage{physics}
\usepackage{color}
\usepackage{ulem}
\usepackage[utf8]{inputenc}
\usepackage{comment}

\begin{document}
\title{Vortices in Fermi gases with spin-dependent rotation potentials}

\author{T. Ichmoukhamedov}
\email{timour.ichmoukhamedov@uantwerpen.be}
\affiliation{TQC, Universiteit Antwerpen, Universiteitsplein 1, 2610 Antwerpen, Belgium}
\author{J. Tempere}
\affiliation{TQC, Universiteit Antwerpen, Universiteitsplein 1, 2610 Antwerpen, Belgium}
\affiliation{Lyman Laboratory of Physics, Harvard University, Cambridge, Massachusetts 02138, USA}
\date{\today}

\begin{abstract}
The rotation of two-component Fermi gases and the subsequent appearance of vortices have been the subject of numerous experimental and theoretical studies. 
Recent experimental advances in hyperfine state-dependent potentials and highly degenerate heteronuclear Fermi gases suggest that it would be feasible to create component-dependent rotation potentials in future experiments. 
In this study we use an effective field theory for Fermi gases to consider the effects of rotating only one component of the Fermi gas. We find that the superfluid band gap in bulk exists up to higher rotation frequencies because the superfluid at rest, far away from the vortex, has to resist only half of the rotational effects.
The vortex remains the energetically favorable state above a critical frequency but exhibits a larger core size. 
\end{abstract}


\maketitle

\section{INTRODUCTION}
Superfluid Fermi gases have recently become of great interest in the field of ultracold atomic physics because of the many modifiable experimental parameters related to the fermionic substructure of the correlation pairs in the condensate, allowing to probe
pairing in regimes inaccessible in solids.
The interaction strength between different components of the Fermi gas can be tuned using Feshbach resonances \cite{FeshBachReview}, granting experimental access to superfluid regimes across the BCS-BEC crossover \cite{Jin,Thomas,Salomon,Grimm1}. 
Subsequently, superfluidity in the presence of population imbalance between the pairing  partners was realized experimentally \cite{ZwierleinImbalance,PartridgeImbalance}, which led to renewed theoretical investigations of spin imbalance effects \cite{ImbalancedFermiGasesReview}. 
Also spin-orbit coupling \cite{FermiSOCWang} 
as well as dimensionality \cite{Fermi2DFeld}
can be tuned. Even greater control of the individual spin components has been achieved by using magnetic field gradients to create spin-dependent potentials \cite{SpinDependent1,SpinDependent2} and in another approach (thusfar limited to Bose gases) hyperfine state-dependent microwave potentials have been created \cite{Bohi2009,Riedel2010}. 

Furthermore, recent observations of degenerate Fermi mixtures of dysprosium and potassium \cite{GrimmMixture} are a first step towards the superfluid state in Fermi gases with components of different atomic species, which would grant unprecedented control over the individual components of the gas. 
This naturally raises the question as to what novel phenomena could be revealed by using state-dependent potentials in experiments that previously relied on simultaneous control of both components of the condensate. In this paper, we answer this question in the context of vortex formation in a rotated fermionic superfluid. 

Vortices, one of the hallmarks of superconductivity and superfluidity, have been created in Fermi gases \cite{FermiVortex, ZwierleinImbalance,FermiVortex2} by stirring the condensate with detuned lasers. 
Although the effect of a spin-dependent population (spin-imbalance) was investigated experimentally in a rotated Fermi superfluid \cite{ZwierleinImbalance}, both spin components were subject to the same rotational potential. 
Other techniques such as phase imprinting have also been used to achieve vortices in Bose Einstein condensates (BECs) \cite{PhaseImprintingBEC} and phase imprinting individual components of Fermi gases has been proposed for soliton creation \cite{PhaseImprintingSoliton}. 
Rotation and vortices in Fermi gases where both components are subject to rotation have been the subject of extensive theoretical study \cite{VortexBruun,StringariVortex,BDG,RotationUrban,VortexWarringa1,VortexWarringa2,
Simonucci2015,KliminVortex1,KliminVortex2}, however here we consider the rotation of both components at different frequencies. 

To study this, we introduce spin-dependent rotation potentials in the microscopic action of the Fermi gas and consider the effects of rotating only a single component of the gas on the bulk value of the superfluid gap and on the first quantized vortex state. 
Our approach is based on the effective field theory (EFT) for superfluid Fermi gases presented in \cite{EFTOriginal,EFTsoliton,EFTSnake} which has also been applied to the study of vortices \cite{KliminVortex1,KliminVortex2}. 
Introduction of spin-dependent vector potentials for the two components in the 
microscopic action leads to a modification of the local chemical potential 
and spin imbalance parameter and yields additional terms in the EFT action.
We solve the modified gap and number equations to study the uniform background amplitude of the superfluid gap of the system away from the disturbance. 
Finally we use a variational approach \cite{VERHELST201796} to probe the critical rotation frequencies for vortex creation and the behavior of the vortex core size.

\section{Spin-dependent rotation in the effective field theory.}

This section presents an overview of our approach, see also appendix (\ref{Appendix A}) and \cite{EFTOriginal,EFTsoliton,EFTSnake,KliminVortex1,KliminVortex2} for details. 
For a Fermi gas of particles with mass $m$ and density $n_0$ the quantities in the rest of this work will be given in units of the Fermi vector $k_F=(3 \pi^2 n_0)^{1/3}$, the corresponding Fermi energy $E_F=\hbar^2 k_F^2 / 2m$ and the Boltzmann constant $k_B$.
The two spin component ($\sigma=(\uparrow,\downarrow)$) Fermi gas is described
by Grassmann fields $(\psi_{\uparrow}$, $\psi_{\downarrow})$ that interact through contact interactions with strength $g$, and that are confined by a trap with potential $V(\mathbf{x})$. The Lagrangian density of this system is given by:
\begin{widetext}
\begin{align}
  \mathcal{L}=  \sum_{\sigma} \bar{\psi}_{\mathbf{x},\tau,\sigma} \left(\frac{\partial}{\partial \tau} - \left( \nabla_{\mathbf{x}} - i\mathbf{A}_\sigma(\mathbf{x}) \right)^2   + V_{\sigma}(\mathbf{x}) - \mu_{\sigma} \right) \psi_{\mathbf{x},\tau,\sigma} \
+  g \bar{\psi}_{\mathbf{x},\tau,\uparrow}\bar{\psi}_{\mathbf{x},\tau,\downarrow} \psi_{\mathbf{x},\tau,\downarrow}\psi_{\mathbf{x},\tau,\uparrow}.
\label{LagrangianDensity}
\end{align}
\end{widetext}
The rotation potentials are time-dependent in the lab frame, but can be represented as time-independent vector potentials $\mathbf{A_{\sigma}}(\mathbf{r})$ through a transformation into the rotating frame of reference. 
This standard procedure corresponds to replacing the gradient for the spin $\sigma$ component by $\nabla_{\mathbf{x}}  \rightarrow \nabla_{\mathbf{x}}  - i\mathbf{A}_\sigma(\mathbf{x}) $ and the trapping potential by $V_{\sigma}(\mathbf{r})= V(\mathbf{r}) - \mathbf{A_{\sigma}}(\mathbf{r})^2$. We want to emphasize that when fully expanded, the Lagrangian contains no quadratic term in the vector potential, and (\ref{LagrangianDensity}) is merely a convenient notation. The vector potentials are written as $\mathbf{A_{\sigma}}=\frac{1}{2} \boldsymbol{\omega_{\sigma}} \times \mathbf{r}$
where $\boldsymbol{\omega_{\sigma}}$ is the spin-dependent rotation vector (whose length equals the rotation frequency $\omega_{\sigma}$ for spin $\sigma$, and whose orientation indicates the axis and sense of rotation).

In the following we will consider a uniform box potential and set $V(\mathbf{r})=0$, where the hard walls of the potential are implemented through a self consistent normalization condition in the trapping region. While in many of the experiments with Fermi gases harmonic traps are used, uniform potentials are no longer an unfeasible setup to consider. This choice is motivated by a number of recent experimental breakthroughs where uniform potentials with hard walls have been realized in BECs \cite{HomogeneousTrapBose}, in 2D Fermi gases \cite{2DFermiBoxPotential}, in 3D Fermi gases \cite{HomogeneousTrapFermi,Baird2019}, and more recently even a 3D cylindrical configuration has been achieved \cite{Zwierlein2019}.

To obtain an effective field theory corresponding to Lagrangian (\ref{LagrangianDensity}), first a bosonic pair field $\Psi(\mathbf{r},t)$ is introduced through the Hubbard-Stratonovic transformation to decouple the quartic interaction term. Then, after performing the fermionic path integrals, the effective action for the pair field is expanded up to second order in the gradients. 
The expansion is valid under the assumption that the bosonic field varies sufficiently slowly, i.e.~on length scales larger than the pair correlation length $\xi_p$ and energy scales smaller than $2\Delta$, the spectroscopic pair-breaking gap.
As shown in \cite{EFTsoliton}, the applicability domain of the EFT comprises the BEC-BCS crossover, except for the BCS regime at low temperatures.
The spatial dependence of the potentials is taken into account in the local density approximation (LDA) which makes the additional assumption that significant changes in $\mathbf{A}_{\sigma}(\mathbf{r})$ take place at larger scales than $\xi_p$. 
Accordingly, we perform the effective action expansion up to $|\nabla \Psi|^2$ in the kinetic energy terms and in addition neglect gradients in the rotational terms beyond the linear contribution $ \nabla \Psi$.
 This is to obtain the same order of expansion on the two sides of the BCS-BEC crossover as the Ginzburg-Landau BCS or Gross-Pitaevskii BEC action functionals. 
The resulting effective action yields the free energy functional (time independent):
\begin{align}
F=&\int \dd \mathbf{r} \text{\hspace{3pt}} 
  \bigg\{ \Omega_s(\mathbf{r}) + C(\mathbf{r}) \left| \nabla \Psi \right|^2 
  - E(\mathbf{r}) ( \nabla |\Psi|^2)^2  \nonumber \\
& + i \left[ D(\mathbf{r}) \left(\frac{\mathbf{A _{\uparrow} + A_{\downarrow}}}{2} \right) 
  + U(\mathbf{r}) \left(\frac{\mathbf{  A_{\downarrow}-A _{\uparrow}}}{2} \right) \right]  
   \cdot  \left( \Psi^* \nabla \Psi - \Psi \nabla \Psi^* \right) \bigg\}.
\label{FreeEnergy}
\end{align}
The free energy (\ref{FreeEnergy}) acquires a term that contains the difference of the vector potentials $\mathbf{  A_{\downarrow}-A _{\uparrow}}$, and that is only present in the case of spin-dependent rotation. 
Expressions for the coefficients $C(\mathbf{r})$,  $E(\mathbf{r})$, $D(\mathbf{r})$, $U(\mathbf{r})$ and $\Omega(\mathbf{r})$ can be found in appendix (\ref{Appendix A}). 
To calculate these coefficients, one needs the values of the 
chemical potentials and 
the superfluid pair breaking gap $\Delta_{\infty}$
for the uniform case (i.e.~without vortex). 
These can be related to the densities and interaction strength 
using the gap equation and equation of state deemed most appropriate, 
for example the experimentally determined equation of state. 
Here, we use the mean-field equation of state as we want to focus 
more on the qualitative results than on a quantitative comparison.

It will prove to be illustrative to consider expression (\ref{FreeEnergy}) from two separate viewpoints.
While the gradients of the field $\Psi$ in expression (\ref{FreeEnergy}) describe the center-of-mass motion of the bosonic pairs, the contributions due to the internal structure 
of the pairs are taken into account in the coefficients. 
At this level, the rotational effects are shown to be present through position and momentum dependent extensions of the bulk chemical potential and spin imbalance:
\begin{align}
\zeta_{\mathbf{k}}(\mathbf{r})=\zeta + \mathbf{k} \cdot (\mathbf{ A _{\uparrow}(\mathbf{r}) + A_{\downarrow}}(\mathbf{r})) , \label{spinimba}\\
\mu_{\mathbf{k}}(\mathbf{r})=\mu +\mathbf{k} \cdot (\mathbf{ A _{\uparrow}(\mathbf{r})- A_{\downarrow}}(\mathbf{r})), \label{chempot} 
\end{align}
where $\mu=(\mu_{\uparrow}+\mu_{\downarrow})/2$ and $\zeta=(\mu_{\uparrow}-\mu_{\downarrow})/2$. The local spin imbalance $\zeta_{\mathbf{k}}(\mathbf{r})$ and chemical potential $\mu_{\mathbf{k}}(\mathbf{r})$ appear in all the coefficients, but their 
effect is most pronounced 
in the zeroth-order $\Omega_{s}(\mathbf{r})$ term which only depends on the local pair density, $\vert \Psi(\mathbf{r}) \vert^2$, and describes the thermodynamic potential of the pair condensate at rest.
To gain a  physical understanding of the modified expressions it is therefore sufficient to consider a stationary superfluid. 
Moreover, as a first approximation (see also comments in the next subsection on the rigidly rotating broken pairs) the superfluid part of the gas is in general indeed stationary when the gas is rotated below the critical vortex frequency, assuming no stirring anisotropy \cite{StringariVortex}.

The local spin imbalance, given by expression (\ref{spinimba}), can be understood by considering the effect of the LDA within the EFT. 
Within this approximation the gas is subdivided into separate elements (larger than the pair correlation length) between which the effects of rotation are uncorrelated.
Each of these elements can now be assigned its own Fermi sphere corresponding to the local density. 
Spin-dependent rotation will induce a displacement in the Fermi spheres of the two components along the local velocity of rotation $\boldsymbol{\omega_{\sigma}} \times \mathbf{r}$.
When the Fermi spheres of both spin components are displaced, pair formation 
is suppressed as the momentum distribution of fermions is no longer symmetric around zero momentum. 
This is akin to the suppression of pair formation due to spin imbalance. 
Indeed, pair formation is strongest at the Fermi surface, 
but in our case the Fermi surfaces are displaced such that in the momentum direction corresponding to that of the rotation velocity, fermions at one Fermi surface cannot find opposite spin, opposite momentum pairing partners.
The modified spin imbalance parameter $\zeta_{\mathbf{k}}$ therefore describes pair breaking due to rotation. 
Note that when the vector potentials are set equal, the expression reduces to $\zeta_{\mathbf{k}}(\mathbf{r})=\zeta + 2\mathbf{k} \cdot \mathbf{ A }(\mathbf{r})$, in accordance with \cite{KliminVortex2}.

To understand the significance of the modified chemical potential (\ref{chempot}) in our approach it will prove to be useful to first consider 
the case of a non-interacting Fermi gas of which the two components are being rotated simultaneously.
From a local density approximation viewpoint, at the microscopic level this rotation is reflected as the aforementioned shift of the local Fermi spheres of each component along the rigid body rotation velocity. 
Each of the two Fermi spheres can now be partitioned in a volume of states that are still symmetric about $\mathbf{k}=0$ and a shifted part that will carry the rotational effects. 
If the interactions between the components were now to be switched on, it is precisely these former symmetric parts of the available states that will form the superfluid at rest.
Here, we assume that all fermions below the Fermi surface participate equally in superfluid formation, which is sufficient for the purpose of this discussion and valid in the BEC regime.
The superfluid in this case exhibits a certain symmetry with respect to exchanging the spins, i.e. for every state ($\mathbf{k}$, $\uparrow$) that is paired with ($\mathbf{-k}$, $\downarrow$) there will also be a pair with the spins interchanged, and hence the average momentum of each component participating in pair formation vanishes. 
The states in the remaining shifted part of the Fermi spheres have no partner of opposite momentum and will remain unpaired and form the normal component that carries the rotational effects. 
This description is closely related to the phase separation between the two components studied in \cite{PhaseSeparation1,PhaseSeparation2,PhaseSeparation3}, where it has been suggested that in the absence of vortices the Fermi gas will form a stationary superfluid in the center with a rapidly rotating normal component that is pushed outwards.
This phenomenon, which requires a complete description of the normal component, is however not captured within our approach and we will only focus on describing the superfluid component in this work.

The picture discussed above changes drastically when only the spin-up component is subject to rotation. In this case the superfluid part can still be formed at rest, but now the spin interchange symmetry is broken. 
Momentum states where $\mathbf{k}$ is aligned with the Fermi sphere shift, contain more pairings of ($\mathbf{k}$, $\uparrow$) with ($-\mathbf{k}$, $\downarrow$) than the case with interchanged spins. 
This means that the average momentum of each of the components participating in pair formation is no longer zero in the case of single-component rotation. 
The superfluid part, even though it is at rest, now consists of two components that each have an average total momentum in opposite directions. 
This means that rotational effects should be directly present in the description of the superfluid component, contrary to the aforementioned situation where two components are rotated simultaneously.
Accordingly, in our description within the LDA, rotational effects appear at the level of the superfluid as a Thomas-Fermi potential in expression (\ref{chempot}). 
Therefore we expect the reappearance of the centrifugal-like force at the level of the superfluid component in our approach when $\mathbf{A_{\uparrow}} \neq \mathbf{A_{\downarrow}}$.
We want to emphasize that this is unrelated to the rotational term $\sim \nabla \Psi$ in the free energy, which is related to the force due to the center of mass motion of the pairs and will lead to vortex creation.

While in this work we will restrict ourselves to studying the single-component rotation case, we want to add a brief  discussion of the case where the frequencies are unequal but neither is zero. In this case, the pair-breaking strength which acts similar to spin imbalance, is determined by the average of the two vector potentials as can be seen in (\ref{spinimba}). Similarly, the previously discussed centrifugal-like force on the non-rotating superfluid component, is determined by the difference of the two in (\ref{chempot}). The discussion presented for the single-component rotation case can now straight forwardly be generalized. At the LDA level, both Fermi spheres will now shift according to their respective local vector potential. This results in a fraction of fermions without a partner of opposite spin, being a measure of the pair-breaking effect. The superfluid part of fermions that could be paired this way, consists of two components with a non-zero local average momentum, resulting in a centrifugal-like force on the non-rotating superfluid part. We can immediately see that one particular interesting case is the one with opposite frequencies $\mathbf{A_{\uparrow}}= \mathbf{A_{\downarrow}}$ where the pair-breaking effect would be minimal and the centrifugal-like force maximal. This seems to signal the necessity to consider arguments beyond the LDA and mean-field for these more exotic cases. 
 
\section{Results}

\subsection{Superfluid gap background amplitude}

In this subsection first the case of a rotated superfluid without vortex is considered, and a comparison is made between the background amplitude of the superfluid order parameter for spin-dependent single-component rotation ($\boldsymbol{\omega}=\boldsymbol{\omega}_{\mathbf{\uparrow}}, \hspace{3pt} \boldsymbol{\omega}_{\mathbf{\downarrow}}=0 $) on the one hand and that for two-component rotation ($\boldsymbol{\omega}=\boldsymbol{\omega}_{\mathbf{\uparrow}}=\boldsymbol{\omega}_{\mathbf{\downarrow}}$) on the other hand. 
The background amplitude is obtained by neglecting all the gradients of the bosonic field and minimizing the lowest-order term of the free energy (\ref{FreeEnergy}) as a function of a uniform superfluid pair breaking gap $\vert \Psi \vert=\Delta_{\infty}$ while demanding particle number conservation. 
For a vortex state the resulting gap yields the bulk solution far away from the center.

\begin{figure}
\includegraphics[width=0.8\columnwidth]{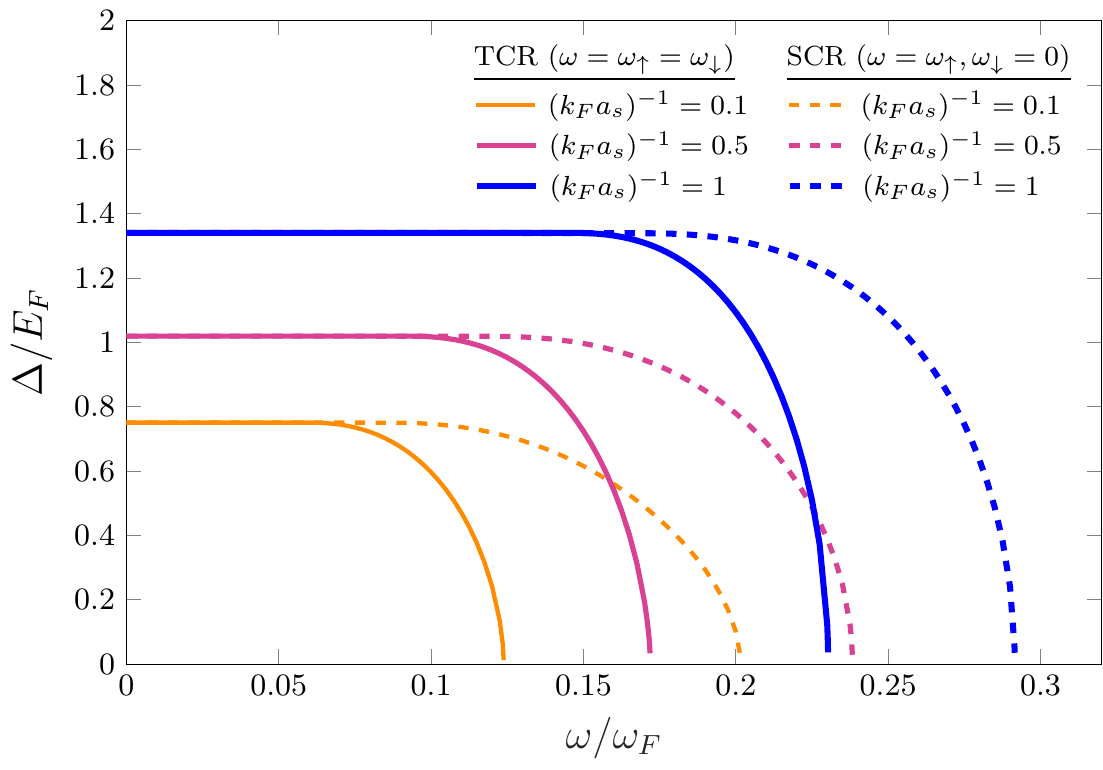}
\caption{The uniform background superfluid gap amplitude $\Delta$ as a function of the applied rotation frequency for two-component rotation (TCR) (solid lines) and single-component rotation (SCR) (dashed lines) for different values of $(k_F a_s)^{-1}$= [0.1, 0.5, 1], respectively using orange, magenta and blue lines (light gray to dark gray). } 
\label{DeltaPlot1}
\end{figure}

The minimization of the free energy results in the mean-field gap and number equations for ($\Delta_{\infty},\mu$) which now contain the spin-dependent rotational effects through the presence of the modified expressions (\ref{spinimba}) and (\ref{chempot}) in $\Omega_s$. 
To solve the equations we consider a Fermi gas in an infinitely long cylindrical container confined by a hard-wall potential. 
The container radius is taken to be $k_F R=15$ in the rest of this work.  Considering the aforementioned EFT region of validity in combination with the LDA, we restrict our study to $(k_F a_s)^{-1} \ge 0.1$ on the BEC side close to unitarity. In addition, as has also been shown in other works \cite{KliminVortex1}, the stability domain of the single-vortex state (which is the model system we will use further on) quickly vanishes on the BCS side further from unitarity. 
To speed up the numerical calculations a momentum cutoff of $\Lambda = 140 k_F$ is used, the length scale of which corresponds to $\Lambda^{-1} \approx 2.1$ nm for the parameters of the ${}^{6}$Li vortex experiment \cite{FermiVortex}. 
This value is chosen to coincide with the estimated range of the interatomic potential for Li - Li interactions \cite{LithiumRangeWerner} and we checked that our results do not significantly change by increasing $\Lambda$ even further. Finally, for numerical convenience we work at a finite temperature corresponding to $T=0.01$ $T_F$, which is almost an order of magnitude lower than in \cite{FermiVortex} and for the purposes of Fig.~(\ref{DeltaPlot1}) represents zero temperature. 

The results presented in Fig.~(\ref{DeltaPlot1}) are in qualitative agreement with the discussion in the previous subsection.
The displacement of the local Fermi spheres, reflected in the $\mathbf{k}$-dependent part of $\zeta_{\mathbf{k}}$, decreases the number of available boson pairs at rest, which adversely affects superfluidity and completely destroys it at a critical value $\omega_{\textrm{max}}$. 
Qualitatively, this behavior is in agreement with the presented values for the asymptotic $\Delta_{\infty}$ in \cite{Simonucci2015}, where its value changes only a little at low rotational frequencies but then shows signs of a rapid decrease as the critical rotation frequency is approached. 
In the present work a homogeneous trapping potential is considered, while the authors of \cite{Simonucci2015} utilize a harmonic trap with a much larger Thomas-Fermi radius. 
A quantitative comparison is therefore not possible since the observed critical values are all much larger in the present case of a small trap.
 
For the rotation of a single component only one of the Fermi spheres undergoes a displacement. 
This increases the number of available pairing states and as a result superfluidity persists up to higher rotation frequencies. 
However, the critical frequencies of single-component rotation do not precisely equal twice the critical frequencies of two-component rotation as would be expected from this reasoning.
The reason for this is the additional adverse effect of $\mu_{\mathbf{k}}$ on the superfluid gap.
As mentioned previously, the modification of the chemical potential can be interpreted as an additional centrifugal force on the condensate. 
This favors pair formation towards the edge, where the pair breaking effect of $\zeta_{\mathbf{k}}$ is larger, and hence results in a detrimental effect on the background amplitude of the gap. 

\subsection{Singly quantized vortex state}

In this subsection a comparison of the lower and upper critical frequencies of the vortex state is presented and the behavior of the vortex core size for two-component rotation and single-component rotation is described. 
It is well known that at rotation frequencies above a critical value, the singly quantized vortex is the energetically favored state \cite{VortexWarringa1}. At even higher frequencies, a vortex lattice will appear \cite{Lattice1Feder,Lattice2Tonini}, or a multiply quantized vortex may be formed \cite{FetterPRA71}. 
The main goal of the present work is to consider the behavior of the singly quantized vortex, which is relevant for the case of frequencies not much larger than the critical rotation frequency, i.e.~in the regime where the density of vortices in the lattice is still small. 
To provide a lower bound estimate of this regime for this system, a two vortex solution will be considered as well, yet for completeness we present and discuss our results for the single-vortex state across the entire frequency regime at which the superfluid exists.
We consider the full free energy in expression (\ref{FreeEnergy}) and subtract its value for the previously obtained uniform solution at rest $F \left[ \Delta_{\infty} \right]$ from the free energy of the rotating vortex state $F \left[ \Psi_\textrm{v}  \right]$:
\begin{equation}
\delta F = \left( F \left[ \Psi_\textrm{v}  \right] - F \left[ \Delta_{\infty} \right] \right) / (2 \pi H),
\label{FreeEnergyDifference}
\end{equation}
where H is the container height. The change in sign from positive to negative $\delta F$ as a function of $\omega$ determines the  critical transition frequency to the vortex state $\omega^{(1)}_c$. 
As $\omega$ is increased more, $\delta F$ changes sign again at the upper critical transition frequency $ \omega^{(2)}_c$ and the system re-enters the superfluid state without vortex. 
Note that $ \omega^{(2)}_c$ can in general lie below the critical pair-breaking frequency $\omega_{\textrm{max}}$ above which the system becomes normal, as has been shown for two-component rotation of harmonically confined Fermi gases \cite{VortexWarringa1,KliminVortex1}.
The vortex state in a cylindrical container with coordinates $(r,\theta)$ and volume $V$ is described by a variational ansatz:
\begin{equation}
\Psi_{\textrm{v}}(\mathbf{r}) =  
   \frac{ \Delta_{\infty} \sqrt{V}}
   { \sqrt{ \int \tanh^2(r/\xi_\textrm{v} ) \dd V }} 
   \tanh \left( \frac{r}{\xi_\textrm{v} } \right) e^{il\theta},
\label{ansatz}
\end{equation}
where $\xi_{\textrm{v}}$ is a variational parameter and $l$ is the vortex winding number ($l=1$ for the singly quantized vortex in this work). 
For a vortex in a superfluid Fermi gas the hyperbolic tangent has been shown to be an excellent description of the vortex core size ($ \leq 1 \%$ error) and to provide a good estimate ($< 1 \%$ error) of the vortex state energy within the EFT, away from the highly spin-imbalanced limit \cite{VERHELST201796}. Note that this does not necessarily imply that (\ref{ansatz}) is an accurate description of the vortex profile itself (as the relative deviations of the profile can be larger), but provides accurate estimates for the core size and energy, which will be our quantities of interest in what follows. 
The superfluid pair breaking gap background $\Delta_{\infty}$ and corresponding chemical potential $\mu$ are obtained by solving the mean-field gap and number equations in the previous section. 
The confinement in the hard-wall box is self-consistently imposed through the normalization condition in (\ref{ansatz}), keeping the total number of pairs fixed in the cylinder of radius $R$. This neglects the healing of the pair condensate over a distance $\xi_{\textrm{v}}$ inside of the hard wall at $r=R$, which does not affect the current results as long as $R \gg \xi_{\textrm{v}}$.
Keeping in mind the normalization, we define here a vortex core size $R_{\textrm{v}}$ through $\Psi_\textrm{v} (R_{\textrm{v}} ) / \Delta_{\infty} = \tanh(1)$ for the optimized variational solution, rather than setting $R_{\textrm{v}}  = \xi_{\textrm{v}}$. 

To estimate the frequency at which a second vortex will enter the trap, in addition to the previous variational ansatz a two-vortex state is considered:
\begin{equation}
\Psi(\mathbf{r}) =  
   \frac{ \Delta_{\infty} \sqrt{V} }
   { \sqrt{ \int f_1^2(\mathbf{r}) f_2^2 (\mathbf{r}) \dd V }} f_1(\mathbf{r}) f_2(\mathbf{r}) e^{i l \left[ \theta_1(\mathbf{r}) + \theta_2(\mathbf{r}) \right]} . 
\label{ansatz2}
\end{equation}
Here, $f_1$ and $f_2$ form the hyperbolic tangent single-vortex profiles separated at some distance $L$ which serves as an additional variational parameter, and $\theta_i$ are the respective angles relative to the vortex centers. Note that this formulation does not incorporate mirror charge vortices, which would be necessary to accurately describe the vortex behavior at the edges, but does implicitly incorporate the presence of the container walls through the normalization condition.

\begin{figure*}
\includegraphics[width=\textwidth]{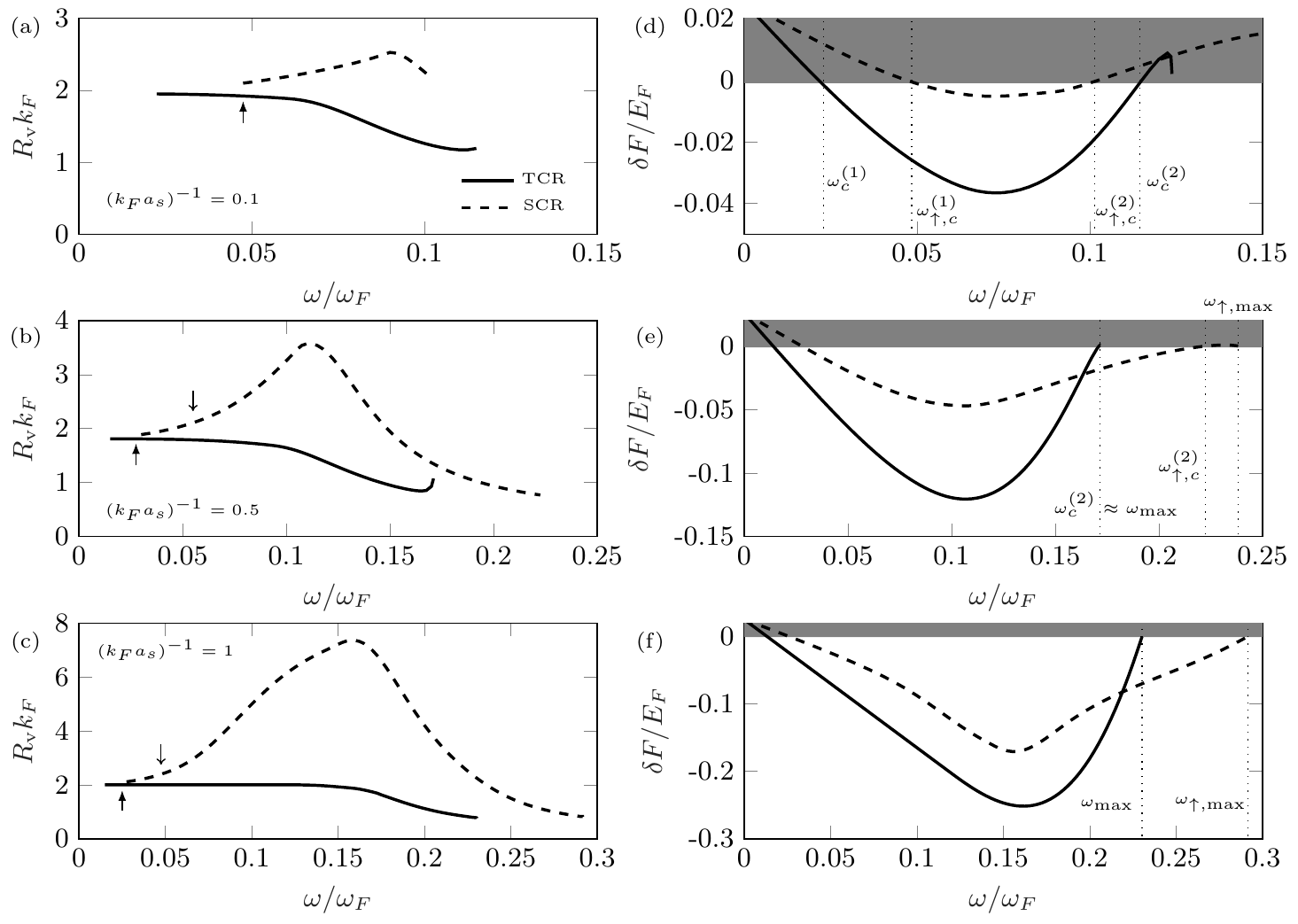}
\caption{The vortex core size $R_{\textrm{v}}$ is shown as a function of the applied rotation frequency for two-component rotation (TCR) (solid line) and single-component rotation (SCR) (dashed line) for $(k_F a_s)^{-1}$= [0.1, 0.5, 1] in (a), (b), (c) respectively, with the corresponding free energy difference from the non-vortex state shown in (d), (e), (f). In the shaded region the free energy is higher than that of the superfluid state without vortex. The transition frequency into the doubly quantized vortex state is indicated on the figure (arrows pointing to the respective SCR and TCR plots). Note that no transition takes place for SCR in (a).} 
\label{VortexPlot}
\end{figure*}

Fig.~(\ref{VortexPlot}) presents the behavior of the vortex core size of the single-vortex state together with the corresponding free energy difference $\delta F$ and the critical frequencies $\omega_c^{(1)}$, $\omega_c^{(2)}$ for TCR and $\omega_{\uparrow,c}^{(1)}$, $\omega_{\uparrow,c}^{(2)}$ for SCR.
While the results and discussion presented below consider the single-vortex state, the transition into the two-vortex state is indicated on Fig.~(\ref{VortexPlot}) as well. Due to the relatively small size of the container the doubly quantized vortex, where both vortex profiles coincide, is found to be the favorable state into which the system transitions. Note that for SCR at $(k_F a_s)^{-1}=0.1$, we find that in the system under investigation the two-vortex state does not appear and the single-vortex state remains favorable.

For TCR, after the vortex is created at $\omega_c^{(1)}$, for a broad initial frequency range, the size of the vortex core remains relatively constant in Fig. (\ref{VortexPlot}) (a), (b) and (c). 
This behavior is in excellent qualitative agreement with the core profile of the central vortex in a vortex lattice presented in \cite{Simonucci2015}, where the vortex profile hardly changes as a function of the rotation frequency. 
The apparent absence of the centrifugal force on the vortex state should not come as a surprise for the ideal singly quantized vortex, because the angular momentum of the bosonic field is homogeneously spread throughout the pair condensate in this model.
Furthermore, as mentioned previously, we do not include the rigidly rotating outer normal gas shell that is expected to have a significant bulging effect on the condensate. 
Around the frequency where $\Delta_{\infty}$ starts to decrease (see Fig.~\ref{DeltaPlot1}) the vortex core shrinks. 
Close to unitarity $\omega_c < \omega_{\textrm{max}}$ and the vortex state briefly transitions into the stationary superfluid phase again at high frequencies before the Fermi gas transitions to the non-superfluid state. This is particularly pronounced on Fig. \ref{VortexPlot}  (a) for the SCR vortex state. Within our description of the vortex (6), considering the rather small container size, this transition can be understood by noting that the vortex state pushes more superfluid outwards which has an energetically unfavorable effect. When the pairs are weakly bound and the superfluid is easily destroyed by rotation this effect is expected to be even more significant. In addition, the energy gain by creating a vortex $-\omega L_z$ is only half as large for SCR. This can be seen on Fig. \ref{VortexPlot} (d) where the energy of the vortex SCR state quickly exceeds the energy of the homogeneous superfluid.
From $(k_F a_s)^{-1}=0.5$ onwards, as the pairs become more strongly bound, this re-entrant phase disappears and the vortex state directly transitions into the non-superfluid phase. 
Although these results are in qualitative agreement with \cite{KliminVortex1}, we note that our approach does not incorporate the shrinking superfluid radius which contributes to the re-entrant phase observed in \cite{KliminVortex1}.

Significant differences can be observed for the behavior of the vortex for SCR. 
First of all, the critical frequencies of respectively vortex creation and disappearance $\omega_{\uparrow,c}^{(1),(2)}$ are higher.
The increase of the initial critical frequency $\omega_{\uparrow,c}^{(1)}$ can be intuitively understood by noting that the contribution from the rotating term $-\omega L_z$ in the free energy, which favors vortex formation, decreases by half for SCR. 
Hence higher frequencies are required to provide sufficient rotational energy to the condensate. 
The increase of the upper critical frequency $\omega_{\uparrow,c}^{(2)}$ in Fig. (\ref{VortexPlot}) (b) and (c) is related to the superfluid pair breaking gap background $\Delta_{\infty}$ which, for single-component rotation, decreased at higher frequencies as compared for two-component rotation, as seen in Fig.~\ref{DeltaPlot1}. 
The most striking difference between single and two-component rotation revealed in Fig.~(\ref{VortexPlot}) is the large size of the vortex core for single-component rotation, which grows as the system is brought deeper into the BEC regime. 
This observation is in accordance with our previous discussion regarding the repulsive character of the modified chemical potential $\mu_{\mathbf{k}}$ in expression (\ref{chempot}). 

 To gain an understanding of why the effect of $\mu_{\mathbf{k}}$ is more prominent in the BEC limit in our approach it is instructive to consider the mean field energy density $\Omega_s$ at zero temperature. In the high momentum limit its integrand  reduces to $\sim \left( \xi_{\mathbf{k}} - E_{\mathbf{k}} + \Delta^2/k^2 \right)$, which for a non-rotating gas describes the modification of the fermionic excitation dispersion as a result of pair formation. 
The $\mathbf{k}$-range in which this expression is non-zero can hence be related to the momentum states that participate in forming the superfluid component. 
On the deep BCS side this range is localized roughly around $\mathbf{k} \approx \mathbf{k_F}$, as expected.
However, moving to the BEC side, this expression obtains extremely long momentum tails indicative of the broadening pair formation window around the Fermi surface. 
Indeed, it is well known that the relative momentum distribution of the fermions in the BEC limit is significantly broadened \cite{MomentumDistribution1,MomentumDistribution2}. To make the connection with SCR, it is sufficient to notice that in our LDA approach the repulsive modified chemical potential $\mu_{\mathbf{k}}$ enters both $E_{\mathbf{k}}$ and $\xi_{\mathbf{k}}$. As more states at higher momentum are present in the superfluid component in the BEC limit, the repulsion of $\mu_{\mathbf{k}}$ is consequently expected to already be greater at the mean field level.
\color{black}

\section{Conclusion}
We have studied the behavior of a two-component superfluid Fermi gas, where only one of the two components is subject to a rotating potential.
We find that not only does the superfluid state persists for single-component rotation, but also that it exists up to higher rotation frequencies.
Above a critical frequency, at a value larger than for two-component rotation, a vortex state appears which exhibits a much larger vortex core than for two-component rotation.
With increasing frequency, initially the size of the vortex core grows for single-component rotation, whereas it remains initially independent of frequency for two-component rotation. 

The model used in the present work does not include the effects of phase separation and likewise does not consider the formation of a vortex lattice. 
Nevertheless, the observed differences between the two types of rotation considered in this work are already appreciable at frequencies below the transition frequency into the  doubly quantized vortex. 
Therefore, we expect the conclusions regarding the behavior of the single vortex core to be observable. 
We believe that one particularly interesting direction would be to consider this topic in mass-imbalanced superfluids where spin-dependent rotation could be feasibly implemented experimentally.

\begin{acknowledgments}
We acknowledge financial support from  
the Research Foundation-Flanders (FWO - Vlaanderen) Grant No. G.0429.15.N and Grant No. G.0618.20.N, 
and from 
the University Research Fund (BOF) of the University of Antwerp.
T.I. acknowledges support of 
the Research Foundation-Flanders (FWO-Vlaanderen) 
through a PhD Fellowship in Fundamental Research, Project No. 1135519N.
\end{acknowledgments}

\appendix
\section{Effective field theory  \label{Appendix A} }

In this appendix a summarized discussion of the EFT is presented. More details on the EFT can be found in \cite{EFTOriginal,EFTsoliton,EFTSnake,KliminVortex1,KliminVortex2,SnakeInstability2019}. The partition sum of a Fermi gas at temperature $\beta=E_F / k_B T =T_F/T$, with $E_F$ the Fermi energy of a free Fermi gas and $k_B$ the Boltzmann constant, can be written as a path integral of the Euclidian action $S$ of the fermionic fields ($\bar{\psi}_{\mathbf{x},\tau,\sigma},\psi_{\mathbf{x},\tau,\sigma}$):
\begin{widetext}
\begin{align}
\mathcal{Z}=\int \mathcal{D} \psi \exp & \left[   -\int \displaylimits_0^{\beta} d\tau \int d \mathbf{x} \sum_{\sigma} \bar{\psi}_{\mathbf{x},\tau,\sigma} \left(\frac{\partial}{\partial \tau} - \left( \nabla_{\mathbf{x}} - i\mathbf{A}_\sigma(\mathbf{x}) \right)^2  +V_{\sigma}(\mathbf{x}) - \mu_{\sigma}  \right) \psi_{\mathbf{x},\tau,\sigma} \right. \nonumber \\
- & \left. g\int\displaylimits_0^{\beta} d\tau \int d\mathbf{x} \bar{\psi}_{\mathbf{x},\tau,\uparrow}\bar{\psi}_{\mathbf{x},\tau,\downarrow} \psi_{\mathbf{x},\tau,\downarrow}\psi_{\mathbf{x},\tau,\uparrow} \right]=\int \mathcal{D} \psi \exp \left(  {-S[\psi]} \right).
\label{finalunitpartition}
\end{align}
\end{widetext}
Expression (\ref{finalunitpartition}) is given in units of $k_F=\left( 3\pi^2 n_0\right)^{1/3}$, $E_F=\hbar^2 k_F^2/ 2m$ and $k_B$, where $m$ is the fermion mass and $n_0$ the fermion density. 
The s-wave contact interactions between different components of the gas are described by a contact potential with strength $g$ which can be related to the experimentally accessible scattering length $a_s$ through the Lippmann-Schwinger equation. 
Componentwise rotation is included in the spin-dependent vector potentials $\mathbf{A_{\sigma}}=\frac{1}{2} \boldsymbol{\omega_{\sigma}} \times \mathbf{r}$ and the confinement potential $V(\mathbf{r})$ becomes shifted in this representation $V_{\sigma}(\mathbf{r})= V(\mathbf{r}) - \mathbf{A_{\sigma}}(\mathbf{r})^2$. 
The global chemical potentials $\mu_{\sigma}$ allow to conserve particle number and impose spin imbalance. As mentioned in the main text the vector potentials are assumed to be slowly varying compared to the pair correlation length of the bosonic pairs and taken into account in the local density approximation (LDA). 
After performing the Hubbard-Stratonovic transformation and the fermionic path integrals, the partition sum is rewritten as a path integral over the bosonic pair field $\Psi$,
\begin{align}
\mathcal{Z}=\int \mathcal{D} \Psi \exp & \left( - S_{\textrm{eff}}[\Psi] \right),
\label{finalunitpartition1}
\end{align}
where the effective action $S_{\textrm{eff}}$ is given by: 
\begin{equation}
S_{\textrm{eff}}=S_B- \text{Tr} \left(\log \left[-\mathbb{G}^{-1} \right] \right).
\label{Sfinal1a}
\end{equation}
In expression ($\ref{Sfinal1a}$) the bosonic part is equal to $S_B=- (1/ g )\int \dd \tau \int \dd \mathbf{x} \left| \Psi \right|^2 $. The second part contains the full propagator $\mathbb{G}$ which can be expressed as
$ -\mathbb{G}^{-1} = -\mathbb{G}_0^{-1} + \mathbb{F}, $
using the pair field matrix $\mathbb{F}(\mathbf{r},\tau)$: 
\begin{equation}
\mathbb{F}(\mathbf{r},\tau)= \begin{pmatrix}0  & -\Psi_{\mathbf{r},\tau}  \\ -\Psi^*_{\mathbf{r},\tau} &  0 \end{pmatrix},
\label{FF}
\end{equation}
and Nambu-Gorkov propagators $\mathbb{G}_0(\mathbf{k},n)$ in reciprocal space:
\begin{equation}
\mathbb{G}_0(\mathbf{k},n) = \begin{pmatrix} \frac{1}{i\omega_n - \xi_{\mathbf{k}} + \zeta_{\mathbf{k}}} & 0 \\ 0 & \frac{1}{i\omega_n + \xi_{\mathbf{k}} + \zeta_{\mathbf{k}}} \end{pmatrix}.
\label{G0k1}
\end{equation}
Here $\omega_n = (2n+1)\pi/\beta$ is the fermionic Matsubara frequency and $\xi_{\mathbf{k}}(\mathbf{r}) =k^2 - \mu_{\mathbf{k}}(\mathbf{r})$ is the shifted kinetic energy. The chemical potential $\mu = (\mu_{\uparrow} + \mu_{\downarrow}) / 2$ and spin imbalance $\zeta = (\mu_{\uparrow} - \mu_{\downarrow}) / 2$ gain a position and wave-vector dependence as discussed in the main text:  
\begin{align}
\mu_{\mathbf{k}}(\mathbf{r}) & = \mu +\mathbf{k} \cdot (\mathbf{ A _{\uparrow}(\mathbf{r})- A_{\downarrow}}(\mathbf{r})) , \label{chempot1}  \\ 
\zeta_{\mathbf{k}}(\mathbf{r}) & = \zeta + \mathbf{k} \cdot (\mathbf{ A _{\uparrow}(\mathbf{r}) + A_{\downarrow}}(\mathbf{r})) . \label{spinimba1}
\end{align}
The pair field matrix is subsequently expanded up to second order in space and time. For this work, only stationary solutions are of importance and time dependence will be neglected:
\begin{align}
\mathbb{F}(\mathbf{r}  + \Delta \mathbf{r}, \tau + \Delta \tau) = \mathbb{F}(\mathbf{r}, \tau ) +  \mathbf{\nabla} \mathbb{F}  (\mathbf{r},\tau) \cdot \Delta\mathbf{r}  + \frac{1}{2} \sum_{\alpha,\beta} \frac{\partial^2 \mathbb{F}(\mathbf{r},\tau)}{\partial x_{\alpha} \partial x_{\beta} } \Delta x_{\alpha} \Delta x_{\beta}.
\end{align}
The effective action functional is expanded up to second order in the gradients. In the terms where the gradients of the field $\nabla \mathbb{F}(\mathbf{r})$ couple to the rotational potentials $\mathbf{A_{\sigma}}$, only the first order gradient contribution is kept and terms such as $\mathbf{A}_{\sigma} \nabla^2 \mathbb{F}(\mathbf{r})$ are not included. The resulting effective action functional is written as:
\begin{align}
S=&\int \limits_0^\beta \dd \tau \int \dd \mathbf{r} \text{\hspace{3pt}} \Bigg\{ \Omega_s(\mathbf{r}) + C(\mathbf{r})\left| \nabla \Psi \right|^2 - E(\mathbf{r}) ( \nabla w)^2 \nonumber \\
& + i \left[ D(\mathbf{r}) \left(\frac{\mathbf{A _{\uparrow} + A_{\downarrow}}}{2} \right)    
  + U(\mathbf{r}) \left(\frac{\mathbf{  A_{\downarrow}-A _{\uparrow}}}{2} \right) \right]  \cdot \left( \Psi^* \nabla \Psi - \Psi \nabla \Psi^* \right) \Bigg\}.
\label{effectiveaction}
\end{align}
The coefficients in expression (\ref{effectiveaction}) are given by:
\begin{align}
  \Omega_s (\mathbf{r}) = & - \int \frac{ \dd \mathbf{k}} {(2\pi)^3} \bigg( \frac{1}{\beta} \log \left[ 2\cosh(\beta E_{\mathbf{k}}) + 2 \cosh(\beta \zeta_{\mathbf{k}}) \right] - \xi_{\mathbf{k}} - \frac{ \left| \Psi \right|^2 }{2 k^2} \bigg) 
  - \frac{ \left| \Psi \right|^2}{8 \pi k_F a_s}.
  \label{OmegaAppendix}
\end{align}
and
\begin{align}
C(\mathbf{r}) &= \frac{2}{3} \int \frac{\dd \mathbf{k}}{(2\pi)^3}  \text{\hspace{3pt}} k^2 f_2(\beta, E_{\mathbf{k}}, \zeta_{\mathbf{k}}), \label{C} \\
E(\mathbf{r}) &= \frac{4}{3} \int \frac{\dd \mathbf{k}}{(2\pi)^3}  \text{\hspace{3pt}} k^2  \xi_{\mathbf{k}} ^2 f_4(\beta, E_{\mathbf{k}}, \zeta_{\mathbf{k}}), \label{E} \\
D(\mathbf{r}) &= \int \frac{\dd \mathbf{k}}{(2\pi)^3}  \text{\hspace{3pt}} \frac{\xi_{\mathbf{k}}}{\left| \Psi \right|^2} \left[f_1(\beta, \xi_{\mathbf{k}}, \zeta_{\mathbf{k}}) - f_1(\beta, E_{\mathbf{k}}, \zeta_{\mathbf{k}}) \right], \label{D} \\
U(\mathbf{r} )&= \int \frac{\dd \mathbf{k}}{(2\pi)^3}  \text{\hspace{3pt}} \frac{\zeta_{\mathbf{k}}}{\left| \Psi \right|^2} \left[f_1(\beta,  \zeta_{\mathbf{k}},\xi_{\mathbf{k}})-f_1(\beta, \zeta_{\mathbf{k}}, E_{\mathbf{k}})  \right]. \label{P}
\end{align}
The pair-breaking excitation energy is equal to $E_\mathbf{k}=\sqrt{\xi_{\mathbf{k}}+\left| \Psi \right|^2}$ and the functions $f_{p}(\beta,\epsilon,\zeta)$ are defined by:
\begin{equation}
f_{p+1}(\beta,\epsilon,\zeta)=-\frac{1}{2p\epsilon} \frac{\partial f_p(\beta,\epsilon,\zeta)}{\partial \epsilon},
\end{equation}
where
\begin{equation}
f_{1}(\beta,\epsilon,\zeta)= \frac{1}{2x} \frac{ \sinh(\beta \epsilon)}{\cosh(\beta \epsilon) + \cosh(\beta \zeta)}.
\end{equation}
In the notation of appendix (A) in \cite{KliminVortex1} we made the same additional approximation $G=D$ as done here to arrive at (\ref{effectiveaction}) and neglect one additional term arising in the derivation that is expected to be negligible within the applicability range of the EFT. The question could be posed whether the term of the spin-dependent rotation coefficient $U(\mathbf{r})$ could also be neglected for the same reason, since an explicit calculation shows it to be extremely small compared to the $D(\mathbf{r})$ term. While this is a valid concern, this would remove the lowest order contribution of spin-dependent rotation at the level of the bosonic field. For the purpose of illustration we will hence keep this term, but note that its contribution will be very small and the main spin-dependent effects will arise due to the modified $\mu_{\mathbf{k}}$ and $\zeta_{\mathbf{k}}$.

Finally, by treating the pair field $\Psi$ as a classical field we can identify the free energy from expression (\ref{effectiveaction}) as:
\begin{align}
F=&\int \dd \mathbf{r} \text{\hspace{3pt}} 
  \bigg\{ \Omega_s(\mathbf{r}) + C(\mathbf{r}) \left| \nabla \Psi \right|^2 
  - E(\mathbf{r}) ( \nabla |\Psi|^2)^2  \nonumber \\
& + i \left[ D(\mathbf{r}) \left(\frac{\mathbf{A _{\uparrow} + A_{\downarrow}}}{2} \right) 
  + U(\mathbf{r}) \left(\frac{\mathbf{  A_{\downarrow}-A _{\uparrow}}}{2} \right) \right]  
  \cdot  \left( \Psi^* \nabla \Psi - \Psi \nabla \Psi^* \right) \bigg\}.
\label{freenergy11}
\end{align}
The derivation of the contributions beyond the saddle-point in the free energy functional (\ref{freenergy11}) relies on a second order gradient expansion $| \nabla \Psi|^2$ in the kinetic energy and a first order gradient expansion $ \boldsymbol{\omega_{\sigma}} \cdot \nabla \Psi$ in the rotational terms. For a consistent treatment some care should be taken as the coefficients $C(\mathbf{r})$, $E(\mathbf{r})$, $D(\mathbf{r})$, $U(\mathbf{r})$ in front of these terms themselves depend on $\boldsymbol{\omega_{\sigma}}$ and $\Psi(\mathbf{r})$. In the coefficients $C(\mathbf{r})$ and $E(\mathbf{r})$ the bosonic field $\Psi(\mathbf{r})$ is replaced by the uniform background amplitude solution $\Delta_{\infty}$, as using $\Psi(\mathbf{r})$ would provide a gradient contribution to the kinetic energy beyond second order \cite{SnakeInstability2019}. Furthermore, the $\boldsymbol{\omega_{\sigma}}$ dependence effectively corresponds to contributions that are at least of the order of $ \boldsymbol{\omega_{\sigma}} |\nabla \Psi|^2$, which were explicitly neglected during the derivation. For an equal treatment of the higher-order rotational effects in the kinetic energy and rotational terms, $\boldsymbol{\omega_{\sigma}}$ is set to zero in $C(\mathbf{r})$ and $E(\mathbf{r})$. For the very same reason $\Psi(\mathbf{r})$ is replaced by $\Delta_{\infty}$ in the coefficients $D(\mathbf{r})$ and $U(\mathbf{r})$, but the $\boldsymbol{\omega_{\sigma}}$ dependence is kept since the contribution will remain maximally first-order in the gradient of the field.

%

\end{document}